\documentclass[wsdraft]{ws-rv9x6} 
\usepackage{subfigure}   
\usepackage{ws-rv-thm}   
\usepackage{ws-rv-van}   

\usepackage{amssymb}
\usepackage{bm}

\newcommand{\be}{\begin{eqnarray}}
\newcommand{\ee}{\end{eqnarray}}
\makeindex
\begin{document}

\chapter[Information processing and Fechner's problem as a choice of arithmetic]{Information processing and Fechner's problem as a choice of arithmetic}\label{ra_ch1}

\author[M. Czachor]{Marek Czachor}
\address{Katedra Fizyki Teoretycznej i Informatyki Kwantowej,\\
Politechnika Gda\'nska, 80-952 Gda\'nsk, Poland\\
and\\
Centrum Leo Apostel (CLEA),\\
Vrije Universiteit Brussel, 1050 Brussels, Belgium,\\
mczachor@pg.gd.pl}

\begin{abstract}
Fechner's law and its modern generalizations can be regarded as manifestations of alternative forms of arithmetic, coexisting at stimulus and sensation levels. The world of sensations may be thus described by a generalization of the standard mathematical calculus.
\end{abstract}

\body

\section{Introduction}

Human beings operate like information processing devices. Since the influential book by Fechner on `elements of psychophysics' \cite{Fechner} it is known that relations between external stimuli and internal sensations occurring in our brains can be modeled mathematically. The models can be tested experimentally in analogy to, or even by means of physical measurements. Hence the term `psychophysics', coined by Fechner.

Fechner himself was a physicist but psychophysics in general does not attract attention of modern pure physicists. A notable exception seems the work of Norwich \cite{Norwich} on information-theoretic foundations of the laws of perception. The fact that formally psychophysics may share some elements with pure physics was intuitively felt by psychologists already in 1930s \cite{Stevens1935}, who understood that psychological experiments are essentially as operational as quantum measurements \cite{Busch}, while `inner psychophysics' of Fechner is as beyond scientific reach as putative hidden variables in quantum mechanics, or interiors of black holes in general relativity. In this sense the degree of objectivity and repeatibility of results of psychophysical measurements are similar to what one encounters in fundamental physics.

For a pure theoretical physicist the field of psychophysics may be, however, interesting also for other reasons.  The goal of the present paper is to look at psychophysics as a non-trivial, theoretical and experimental example of a natural science where a non-Diophantine arithmetic \cite{Burgin1997,Burgin2010} plays a prominent role.

Non-Diophantine arithmetic is determined by a function $f$ and its inverse $f^{-1}$ (or, more generally, by two independent functions $f$ and $g
$). The role of $f$ is similar to that played in psychophysics by the `Fechner function' \cite{LE}. As shown recently \cite{Czachor}, one can reformulate the laws of physics in terms of a non-Diophantine arithmetic and its corresponding non-Diophantine calculus. The formalism found applications in fractal theory \cite{ACK}, but what is still missing is the law that determines the form of $f$.

In this respect non-Diophantine physics is in a similar situation as psychophysics. We need $f$, but we also need a fundamental law that determines it. This is why all approaches to psychophysics which try to understand the fundamental and general laws that govern $f$ are so intriguing.

On the other hand, one may hope that an abstract theoretical-physics insight into the meaning of $f$ may lead to some new ideas for experimental psychology. Anyway, this is how psychophysics started on 22 October 1850...

\section{Non-Diophantine arithmetic and calculus}

Assume the set $X$ is equipped with generalized arithmetic operations (addition $\oplus$, subtraction $\ominus$, multiplication $\odot$, division $\oslash$), defined by \cite{Czachor}
\be
x\oplus y &=& f^{-1}\big(f(x)+f(y)\big),\label{+}\\
x\ominus y &=& f^{-1}\big(f(x)-f(y)\big),\label{-}\\
x\odot y &=& f^{-1}\big(f(x)f(y)\big),\label{.}\\
x\oslash y &=& f^{-1}\big(f(x)/f(y)\big),\label{/}
\ee
where $x,y\in X$, and $f: \mathbb{R}\supset X\to Y\subset\mathbb{R}$ is a bijection. The set $Y$ is equipped with the `standard' arithmetic of real numbers: $\pm$, $\cdot$, and $/$. Neutral elements of addition and multiplication in $X$ are defined by
\be
0' &=& f^{-1}(0),\\
1' &=& f^{-1}(1),
\ee
since then $0'\oplus x=x$, $1'\odot x=x$, $x\oslash x=1'$ (for $x\neq 0'$) and $x\ominus x=0'$ (for any $x\in X$).

One verifies the standard properties: (1) associativity $(x\oplus y)\oplus z=x\oplus (y\oplus z)$,
$(x\odot y)\odot z=x\odot (y\odot z)$, (2) commutativity $x\oplus y=y\oplus x$, $x\odot y=y\odot x$, (3) distributivity
$(x\oplus y)\odot z=(x\odot z)\oplus (y\odot z)$. This is an example of a non-Diophantine arithmetic in the sense of Burgin \cite{Burgin1997,Burgin2010}.

The well known Weber--Fechner problem \cite{BairdNoma} now can be reformulated as follows: Find a generalized arithmetic such that
$(x+kx)\ominus x$ is independent of $x$. In other words, we have to find $f$ solving
\be
(x+kx)\ominus x = f^{-1}\big(f(x+kx)-f(x)\big)=\delta x,\label{delta x}
\ee
with $x$-independent $\delta x$. Acting with $f$ on both sides of (\ref{delta x}) we get
\be
f(x+kx)-f(x)=f(\delta x)
\ee
which is the standard psychophysical Abel problem \cite{LE} for $f$, with constant Weber fraction $\Delta x/x=k$. The solution is $f(x)=a\ln x+b$, $f^{-1}(x)=e^{(x-b)/a}$, and thus $0'=f^{-1}(0)=e^{-b/a}$, $1'=f^{-1}(1)=e^{(1-b)/a}$. Clearly, $0'\neq 0$ and $1'\neq 1$. This type of difference occurs in physics in the approach of Benioff \cite{Benioff,Benioff2} where $f(x)=px$, $p\neq 0$, and $0'=0$ but $1'=1/p$.

Let us try to understand the logical structure of (\ref{delta x}). We have two real numbers, $x$ and $x'=x+kx$, and we have two ways of subtracting them. $x'\ominus x$ clearly corresponds to the sensation continuum, while $x'-x$ is the `usual' way of subtracting employed at the stimulus side. From the arithmetic perspective one expects that not only subtraction, but also addition, multiplication and division are perceived in some `Fechnerian way'. In principle, any form of change can be perceived in a non-Diophantine way.

The idea, when explored in its full generality, leads from non-Diophantine arithmetic to non-Diophantine calculus \cite{Czachor}. In particular, a derivative of  a function $A:X\to X$, is naturally defined by
\be
\frac{d'A(x)}{d'x}
&=&
\lim_{h\to 0'}\big(A(x\oplus h)\ominus A(x)\big)\oslash h.\label{d'A/d'x}
\ee
The derivative should not be confused with the `usual' one,  defined for functions $B:Y\to Y$,
\be
\frac{dB(y)}{dy}
&=&
\lim_{h\to 0}\big(B(y+h)- B(y)\big)/ h.
\ee
An integral is defined in a way that guarantees the two fundamental laws of calculus,
\be
\frac{d' }{d'x}\int_a^x A(x')\odot d'x' &=& A(x),\label{calc1}\\
\int_a^b \frac{d'A(x')}{d'x'}\odot d'x' &=& A(b)\ominus A(a).\label{calc2}
\ee
With any $A:X\to X$ one can associate the conjugate map $B=f\circ A\circ f^{-1}$, $B:Y\to Y$. Then (\ref{d'A/d'x})--(\ref{calc2})  imply
\be
\frac{d'A(x)}{d'x}
&=&
f^{-1}\left(\frac{dB\big(f(x)\big)}{df(x)}\right),\\
\int_a^b A(x)\odot d'x &=&f^{-1}\left(\int_{f(a)}^{f(b)}B(y)dy\right),
\ee
as one can directly verify from definitions.

In order to appreciate the difference between $d'/d'x$ and $d/dx$ take $f(x)=x^3$ and let $\sin_f x=f^{-1}\big(\sin f(x)\big)=\sqrt[3]{\sin (x^3)}$. Then
\be
\frac{d \sin_f x}{dx}
=
\frac{x^2 \cos (x^3)}{\sin ^{\frac{2}{3}}(x^3)},
\ee
whereas
\be
\frac{d'\sin_f x}{d'x}=\sqrt[3]{\cos (x^3)}=\cos_f x.\label{d'sin}
\ee

The non-Diophantine derivative is easier to compute: One only replaces sin by cos in (\ref{d'sin}), and neither $f$ nor $f^{-1}$ get differentiated. This is why we
do not need any continuity or differentiability assumption about $f$. In fact, $f$ can be as weird as the Cantor function \cite{Czachor,ACK}. The property may be useful since in the psychophysical theory of numbers \cite{B1,B2,B3,B4,B5} the corresponding psychophysical functions have discontinuous first derivatives, and in principle can be discontinuous themselves.

Moreover, even the simple case of a power function $f(x)=x^q$, $q\neq 1$, leads to non-differentiability at 0 of either $f$ or $f^{-1}$. Yet, the result (\ref{d'sin}) shows that this is not a difficulty since neither $f$ nor $f^{-1}$ are differentiated in the course of computing $d'/d'x$. The power function is an important alternative to Fechner's logarithm \cite{Stevens,Marks}, similarly to the unification of logarithm and power,  derived by Norwich \cite{Norwich}. In the present proof-of-principle analysis we restrict the examples to Fechnerian $f$.

\section{Non-Diophantine Fechnerian arithmetic operations}

Let us now find the explicit forms of non-Diophantine arithmetic operations corresponding to the Fechnerian case.

\subsection{Addition}
\be
x\oplus y
&=&
f^{-1}\big(f(x)+f(y)\big)
\\
&=&
e^{(f(x)+f(y)-b)/a}
\\
&=&
e^{(a\ln x+b+a\ln y+b-b)/a}
\\
&=&
e^{\ln x+b/a+\ln y}
\\
&=&
xy e^{b/a}=xy/0'.
\ee
In particular,
\be
x\oplus 0' &=& x 0' e^{b/a}=x  e^{-b/a}e^{b/a}=x.
\ee
\subsection{Subtraction}

Subtraction is the only non-Diophantine arithmetic operation that implicitly occurs in the psychophysics literature.
\be
x\ominus y
&=&
f^{-1}\big(f(x)-f(y)\big)
\\
&=&
e^{(f(x)-f(y)-b)/a}
\\
&=&
e^{(a\ln x+b-a\ln y-b-b)/a}
\\
&=&
e^{\ln x-\ln y-b/a}
\\
&=&
e^{-b/a}x/y=0'x/y.
\ee
In particular,
\be
x\ominus x
&=&
e^{-b/a}x/x=e^{-b/a}=0',\\
(x+kx)\ominus x
&=&
e^{-b/a}(x+kx)/x=e^{-b/a}(1+k).
\ee
Assuming that for a single just noticable difference one should find $(x+kx)\ominus x=1'$ one arrives at
\be
e^{-b/a}(1+k)= 1'=e^{(1-b)/a},
\ee
and thus $1+k=e^{1/a}$, $a=1/\ln (1+k)$. This leads to the known form of solution of Abel's equation for the Fechner problem \cite{LE},
\be
f(x)=\frac{\ln x}{\ln(1+k)}+b,
\ee
where $b$ is an arbitrary constant.

A negative of $x$ is
\be
\ominus x
&=&
0'\ominus x\\
&=&
e^{-b/a}0'/x=e^{-2b/a}/x.
\ee
Let us cross-check,
\be
\ominus x\oplus x
&=&
(\ominus x)x e^{b/a}
\\
&=&
(e^{-2b/a}/x)x e^{b/a}
\\
&=&
e^{-b/a}
=0'.
\ee
Note that although $x>0$ in $f(x)=a\ln x+b$, one nevertheless has a well defined negative number $\ominus x=e^{-2b/a}/x$, which is... positive. The apparent paradox disappears if one realizes that we speak of two different types of negativity, defined with respect to two different choices of arithmetic.

\subsection{Multiplication}
\be
x\odot y
&=&
f^{-1}\big(f(x)f(y)\big)
\\
&=&
e^{(f(x)f(y)-b)/a}
\\
&=&
e^{((a\ln x+b)(a\ln y+b)-b)/a}
\\
&=&
e^{(a^2\ln x\ln y+ ab\ln x+ab\ln y+b^2 -b)/a}
\\
&=&
e^{a\ln x\ln y+ b\ln x+b\ln y+b^2/a -b/a}
\\
&=&
x^{a\ln y}x^by^be^{b(b-1)/a}.
\ee
In particular
\be
x\odot 1'
&=&
x^{a\ln e^{(1-b)/a}}x^be^{b(1-b)/a}e^{b(b-1)/a}
\\
&=&
x^{a(1-b)/a}x^b=x,
\\
x\odot 0'
&=&
x^{a\ln e^{-b/a}}x^be^{b(-b)/a}e^{b(b-1)/a}
\\
&=&
x^{-b}x^be^{-b/a}=e^{-b/a}=0'.
\ee
\subsection{Division}
\be
x\oslash y
&=&
f^{-1}\big(f(x)/f(y)\big)
\\
&=&
e^{(f(x)/f(y)-b)/a}
\\
&=&
e^{((a\ln x+b)/(a\ln y+b)-b)/a}
\\
&=&
e^{(\ln x+b/a)/(a\ln y+b)-b/a}
\\
&=&
e^{\ln x/(a\ln y+b)}e^{(b/a)/(a\ln y+b)}e^{-b/a}
\ee
In particular,
\be
x\oslash x
&=&
e^{(1-b)/a}=1'.
\ee
\subsection{Multiplication as a repeated addition}

Everybody knows that $1+1=2$, or $2+2=4$. In non-Diophantine arithmetic these rules hold as well, but in a subtle form. First of all, let us define
$n'=f^{-1}(n)$, $n\in \mathbb{N}$. Now,
\be
n'\oplus m'
&=&
f^{-1}\big(f(n')+f(m')\big)\\
&=&
f^{-1}\big(n+m\big)=(n+m)'\\
n'\odot m'
&=&
f^{-1}\big(f(n')f(m')\big)\\
&=&
f^{-1}\big(nm\big)=(nm)'.
\ee
Similarly,
\be
n'\odot m'
&=&
f^{-1}\big(nm\big)\\
&=&
f^{-1}\big(\underbrace{m+\dots +m}_{n\rm{ times}}\big)\\
&=&
f^{-1}\big(\underbrace{f(m')+\dots +f(m')}_{n\rm{ times}}\big)\\
&=&
\underbrace{m'\oplus\dots \oplus m'}_{n\rm{ times}}.
\ee
So, $1'\oplus 1'=2'$, $2'\oplus 2'=4'=2'\odot 2'$. This is how it looks at the internal sensation space. At the stimulus level the calculation looks somewhat different,
\be
1'\oplus 1' &=&e^{(1-b)/a}\oplus e^{(1-b)/a}=e^{(1-b)/a}e^{(1-b)/a}e^{b/a}=e^{(2-b)/a}=2',\\
2'\oplus 2' &=&e^{(2-b)/a}\oplus e^{(2-b)/a}=e^{(2-b)/a}e^{(2-b)/a}e^{b/a}=e^{(4-b)/a}=4'.
\ee
Multiplication is more involved,
\be
2'\odot 2' &=&e^{(2-b)/a}\odot e^{(2-b)/a}\\
&=&
e^{a\ln e^{(2-b)/a}\ln e^{(2-b)/a}+ b\ln e^{(2-b)/a}+b\ln e^{(2-b)/a}+b^2/a -b/a}
\\
&=&
e^{(4-b)/a}=4'.
\ee

\section{Non-Diophantine Fechnerian exponential function}

Let us now switch from non-Diophantine arithmetic to calculus.

An important example is provided by the exponential function which, by definition, solves the following problem
\be
\frac{d'A(x)}{d'x} &=& A(x),\\
A(0') &=& 1'.
\ee
The unique solution is $A(x)=f^{-1}\left(e^{f(x)}\right)=\exp_f x$ \cite{Czachor}, and thus
\be
A(x)
&=&
e^{(e^{a\ln x+b}-b)/a}\\
&=&
e^{(e^b e^{\ln x^a}-b)/a}\\
&=&
e^{e^b x^a/a}e^{-b/a}.
\ee
Let us check the initial condition:
\be
A(0') &=& A(e^{-b/a}) =e^{(e^b (e^{-b/a})^a-b)/a}=e^{(1-b)/a}=1'.
\ee
It is an instructive exercise to compute the derivative directly from definition:
\be
\frac{d'A(x)}{d'x}
&=&
\lim_{h\to 0'}\big(A(x\oplus h)\ominus A(x)\big)\oslash h\\
&=&
\lim_{h\to 0'}\big(e^{e^b (x\oplus h)^a/a}e^{-b/a}\ominus e^{e^b x^a/a}e^{-b/a}\big)\oslash h\\
&=&
\lim_{h\to e^{-b/a}}\left(e^{e^b (xhe^{b/a})^a/a}e^{-b/a}\ominus e^{e^b x^a/a}e^{-b/a}\right)\oslash h\\
&=&
\lim_{h\to e^{-b/a}}\left(e^{-b/a}\frac{e^{e^b (xhe^{b/a})^a/a}e^{-b/a}}{e^{e^b x^a/a}e^{-b/a}}\right)\oslash h\\
&=&
\lim_{h\to e^{-b/a}}e^{(\ln e^{-b/a+x^ah^ae^{2b}/a-e^b x^a/a}+b/a)/(a\ln h+b)-b/a}
\\
&=&
\lim_{h\to e^{-b/a}}e^{(-b/a+x^ah^ae^{2b}/a-e^b x^a/a+b/a)/(a\ln h+b)-b/a}
\\
&=&
\lim_{h\to e^{-b/a}}e^{(x^ah^ae^{2b}/a-e^b x^a/a)/(a\ln h+b)-b/a}
\\
&=&
\lim_{c\to b}e^{(x^a(e^{-c/a})^ae^{2b}/a-e^b x^a/a)/(a\ln e^{-c/a}+b)-b/a}
\\
&=&
\lim_{c\to b}e^{\frac{e^{b-c}-1}{b-c}e^b x^a/a-b/a}
\\
&=&
e^{e^b x^a/a-b/a}=A(x),
\ee
which was to be proved.

One further finds that
\be
\exp_f(x\oplus y)=\exp_f x\odot\exp_fy.
\ee
The inverse function
\be
\ln_f x &=& f^{-1}\big(\ln f(x)\big)\\
&=&
e^{(\ln f(x)-b)/a}
\\
&=&
e^{-b/a}f(x)
\\
&=&
e^{-b/a}(a\ln x+b)
\ee
satisfies
\be
\ln_f(x\odot y)=\ln_fx\oplus \ln_f y.
\ee

\section{Final remarks}

The fact that the Abel-equation approach to psychophysics may be regarded as an example of non-Diophantine arithmetic is quite evident. The formula for explicit Fechnerian subtraction,
\be
x\ominus y
&=&
e^{-b/a}x/y,
\ee
shows that a ratio at the stimulus level is directly proportional to a difference in the sensation space. $x\ominus y$ is a natural measure of `subjective dissimilarity', and it satisfies the law of additivity,
\be
(x\ominus y)\oplus (y\ominus z)=x\ominus z,
\ee
as required in more modern approaches to the Fechner problem \cite{DC,Falmagne}. One may also wonder if we have here any obvious counterpart of a just noticable difference. $0'$ is a candidate since it is non-zero at the stimulus level, but it marks the threshold of `non-zero change' in the sensation space.

Much more interesting is the issue if experimental psychologists can make sense of the remaining arithmetic operations, and if the corresponding calculus can find experimental applications. I hope the paper will trigger some research in these directions. From the point of view of theoretical physics the ultimate goal is to find a general law that determines the form of arithmetic. Psychophysical insights might be very helpful here.

\end{document}